\documentstyle[aps,prl,epsf,preprint,tighten,floats]{revtex}
\begin{document}
\draft
\title{Charge-Relaxation and Dwell Time in the 
Fluctuating Admittance of a Chaotic Cavity}
\author{P. W. Brouwer$^{\rm a}$ and M. B\"uttiker$^{\rm b}$}
\address{
  $^{a}$Instituut-Lorentz, University of Leiden, P.O. Box 9506, 
    2300 RA Leiden, The Netherlands\\
  $^{b}$D\'epartement de Physique Th\'eorique, Universit\'e de 
    Gen\`eve, CH-1211 Gen\`eve 4, Switzerland}
\maketitle

\begin{abstract}
  We consider the admittance of a chaotic quantum dot, 
  capacitively coupled to a gate and connected 
  to two electron reservoirs by multichannel 
  ballistic point contacts. For a dot  
  in the regime of weak-localization and 
  universal conductance fluctuations, we calculate   
  the average and variance of the admittance 
  using random-matrix theory. 
  We find that the admittance is governed by 
  two time-scales: the classical admittance depends on the
  $RC$-time $\tau$ of the quantum dot, 
  but the relevant time scale for 
  the weak-localization correction and the 
  admittance fluctuations is the dwell time. 
  An extension of the circular ensemble is used for a 
  statistical description of the energy dependence of 
  the scattering matrix. \smallskip
\end{abstract}

A quantum dot is a small conducting island, formed with the help of gates, 
with a ballistic and chaotic classical dynamics,
and coupled to electron reservoirs by ballistic point contacts. 
The search for signatures of phase-coherent transport 
through chaotic quantum dots focused on the 
d.c.\ conductance \cite{JBS,PEI,PAEI,BarangerMello}.
However, the a.c.\ response is also of 
interest \cite{PAEI,GoparMelloBuettiker,KumarJayannavar}, 
since it probes the charge distribution and its dynamics.
While the d.c.\ conductance is entirely determined by the scattering
properties of the quantum dot, a.c.\ transport 
requires that nearby conductors (gates) are taken into 
account as well \cite{BPT1993,Buettiker1993,CHR}: charges 
may temporarily pile up in the quantum dot, thus interacting 
with the gates through long-range Coulomb forces.

Except for highly transmissive samples \cite{CHR},
the low-frequency dynamics of a mesoscopic conductor
is governed by a charge-relaxation mode or an RC-relaxation time $\tau$.
However, as soon as weak localization \cite{AndersonGLK} 
and universal conductance fluctuations \cite{AltshulerLeeStone} play a role, 
this is no longer a complete picture. In this work, we demonstrate
that the weak localization effects and the a.c.\ conductance
(admittance) fluctuations are primarily governed
by a second time-scale, a dwell time $\tau_{d}$, 
characteristic of the non-interacting system.
The large disparity of these two time-scales ($\tau_{d} \gg \tau$
for a macroscopic quantum dot) dramatically affects the admittance 
and provides a signature that should be readily observed.

In a recent paper, Gopar, Mello, and one of the authors 
studied the capacitance fluctuations of a chaotic quantum dot, 
coupled to the outside world through one point contact 
with a single conducting channel only \cite{GoparMelloBuettiker}.
For the low-frequency fluctuations, weak localization effects are absent
and the double-time scale behavior discussed here does not occur.
In this letter, we calculate the average and variance of the
admittance for the case of a two-probe quantum dot with multichannel 
point contacts. Multichannel contacts are necessary
to be in the regime of
weak localization and universal conductance fluctuations.
Moreover, the presence of two point contacts instead of one
turns out to be essential for the existence of quantum interference
effects 
to leading order in the frequency $\omega$.

\begin{figure}
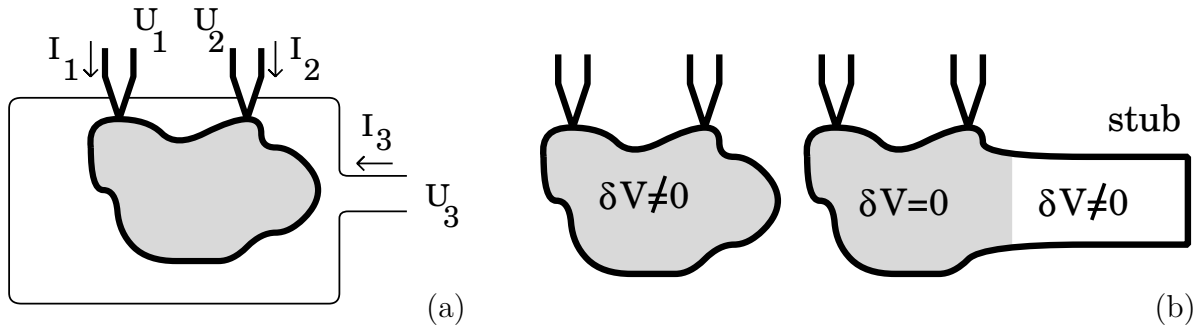


\hspace{0.0\hsize}
\epsfxsize=0.37\hsize
\epsffile{sys.eps}
\vspace{-0.21\hsize}

\hspace{0.05\hsize}
\hspace{0.37\hsize}
\epsfxsize=0.53\hsize
\epsffile{stub.eps}
\par

\hspace{0.34\hsize} (a) \hspace{0.54\hsize} (b) \vspace{0.5cm}

\caption{\label{fig:1} (a) Chaotic cavity (grey), coupled to source 
and drain reservoir (1 and 2) by point contacts. 
The cavity is coupled capacitively to the gate (3). (b)
\label{fig:2} Construction of the energy dependent 
ensemble of scattering matrices. 
A change $\delta \varepsilon$ of the energy is replaced by a spatially 
uniform change $\delta V = -\delta \varepsilon/e$ of the potential 
in the cavity (left), which in turn is statistically equivalent to 
a chaotic cavity with $\delta V = 0$ (right), coupled to a closed 
lead (a stub) with an energy-dependent reflection matrix.}
\end{figure}

The system under consideration is depicted in Fig.\ \ref{fig:1}a. 
Two electron reservoirs at voltages $U_1(\omega)$ and $U_2(\omega)$
are coupled to the quantum dot by two point contacts with
$N_1, N_2 \gg 1$ modes, 
through which currents $I_1(\omega)$ and $I_2(\omega)$ are passed. 
The dot is coupled capacitively to a gate, 
connected to a reservoir at voltage $U_3(\omega)$, 
from which a current $I_3(\omega)$ flows. 
A geometrical capacitance $C$ accounts 
for the capacitive coupling with the gate
\cite{GoparMelloBuettiker,BPT1993}.
We assume that the gate is macroscopic, 
i.e.\ that its density of states $dn_3/d\varepsilon \gg C/e^2$. 
The a.c.\ transport properties of the 
system are characterized by the 
dimensionless admittance 
$G_{\mu \nu}(\omega) = (h/2e^2)\delta I_{\mu}(\omega)/\delta U_{\nu}(\omega)$. 
We restrict ourselves to the 
coefficients $G_{\mu \nu}(\omega)$ with $\mu,\nu=1,2$, 
the remaining coefficients being determined by 
current conservation and gauge invariance \cite{BPT1993,Buettiker1993,CHR},
$
  \sum_{\mu=1}^{3} G_{\mu\nu}(\omega) = \sum_{\nu=1}^{3} G_{\mu\nu}(\omega) = 0.
$
The emittance $E_{\mu \nu}$ is the first order term in a small-$\omega$ 
expansion of the admittance, 
\begin{equation}
  G_{\mu \nu}(\omega) = G_{\mu \nu} - i \omega E_{\mu \nu} + {\cal O}(\omega^2).
\end{equation}
Here $G_{\mu \nu} \equiv G_{\mu \nu}(0)$ is the d.c.\ conductance. 
The emittance coefficients are the analogues of the capacitance 
coefficients for a purely capacitive system \cite{Buettiker1993,CHR}.

A calculation of the admittance proceeds in two steps \cite{BPT1993}. 
First, we calculate the unscreened admittance 
$G^{u}_{\mu\nu}(\omega)$, 
the direct response to the change in the external potentials
(at fixed internal potential)
\begin{eqnarray} \label{eq:Gab1}
  && G_{\mu\nu}^{u}(\omega) =  \int d\varepsilon\,  
  {f(\varepsilon - \case{1}{2} \hbar \omega )-f(\varepsilon 
  + \case{1}{2} \hbar \omega) \over \hbar \omega}
  \mbox{tr}\, 
  \left[\delta_{\mu \nu} \openone_{\mu} 
  - S_{\mu \nu}^{\dagger}(\varepsilon 
  - \case{1}{2} \hbar \omega) S_{\mu \nu}^{\vphantom{\dagger}}(\varepsilon 
  + \case{1}{2} \hbar \omega) \right].
\end{eqnarray}
Here $f(\varepsilon)$ is the Fermi function, 
$S_{\mu \nu}$ is the $N_{\mu} \times N_{\nu}$ 
scattering matrix for scattering from $\nu$ to $\mu$, 
and $\openone_{\mu}$ is the $N_{\mu} \times N_{\mu}$ unit matrix. 
Second, we take the screening due to the long-range Coulomb interactions into 
account, which was ignored in Eq.\ (\ref{eq:Gab1}).
For a single self-consistent potential within the cavity,
the result is \cite{BPT1993}
\begin{eqnarray} \label{eq:GIab1a}
  G_{\mu\nu}(\omega) &=& G_{\mu\nu}^{u}(\omega) +
    {\sum_{\rho=1}^{2} G_{\mu\rho}^{u}(\omega)
    \sum_{\sigma=1\vphantom{\rho}}^{2} G_{\sigma\nu}^{u}(\omega)
    \over  i h \omega C/2e^2 - \sum_{\rho=1}^{2} 
    \sum_{\sigma=1}^{2} G_{\rho\sigma}^{u}(\omega) }.
\end{eqnarray}

The average over the ensemble of quantum dots is performed using 
random-matrix theory \cite{Mehta}. 
We use
an extension of the circular ensemble of uniformly distributed
scattering matrices. This extension provides a statistical description 
of the energy-dependence of the scattering matrix \cite{foot1}.
To construct the extended circular ensemble we
first replace an energy shift $\delta \varepsilon$ 
by a uniform decrease 
$\delta V = -\delta \varepsilon/e$ of the potential $V$ 
in the quantum dot. 
The key point of our method is to localize $\delta V$ in a closed lead
(a stub), see Fig.\ \ref{fig:2}b. 
The stub contains $N_{\rm s} \gg N_1 + N_2$ modes to ensure that it 
models a spatially homogeneous potential drop $\delta V$. 
The system consisting of the dot and the stub is described by 
the $N_{\rm s} \times N_{\rm s}$, $\varepsilon$-dependent 
reflection matrix $r_{\rm s}(\varepsilon)$ of the stub and 
the $(N_1 + N_2 + N_{\rm s})$-dimensional scattering matrix 
$U$ of the cavity at reference energy $\varepsilon_0$,
with the stub replaced by a regular open lead. 
We choose the scattering basis in the stub and the cavity 
such that $r_{\rm s}(\varepsilon_0) = 1$. 
For $\varepsilon$ different from $\varepsilon_0$ we take
\begin{equation} \label{eq:VB}
  r_{\rm s}(\varepsilon) =  e^{i (\varepsilon-\varepsilon_0) \Phi},\ \
  \phi = \mbox{tr}\, \Phi
\end{equation}
where the matrix $\Phi$ is Hermitian and positive definite. 
For $N_{\rm s} \gg N_1 + N_2$, the precise choice of $\Phi$ becomes irrelevant, 
all information being contained in the single parameter $\phi$. 
For the matrix $U$ we assume a uniform distribution.
In the presence of time-reversal symmetry, both $U$ and $\Phi$ are symmetric. 
We finally express the scattering matrix $S(\varepsilon)$ in terms 
of $U$ and $r_{\rm s}(\varepsilon)$,
\begin{eqnarray} \label{eq:CUEB}\label{eq:SU}
  S(\varepsilon) &=& U_{\rm ll} + 
  U_{\rm ls} \left[1 - r_{\rm s}(\varepsilon) U_{\rm ss} \right]^{-1} 
  r_{\rm s}(\varepsilon) U_{\rm sl}.
\end{eqnarray}
The matrices $U_{ij}$ in Eq.\ (\ref{eq:SU}) are the four blocks of $U$, 
describing transmission and reflection from and to the stub (s) or the 
two leads (l).
The parameter $\phi$ is related to the mean level density 
$\langle dn/d\varepsilon \rangle$ via 
$\phi = 2 \pi \langle dn/d\varepsilon \rangle$.

We are now ready to calculate the average and fluctuations of the admittance.
We first compute the average of the unscreened admittance
$G_{\mu \nu}^{u}(\omega)$ with the help of
the diagrammatic technique of Ref.\ \onlinecite{BrouwerBeenakker1996a},
\begin{eqnarray} \label{eq:Tabavg}
  \langle G_{\mu\nu}^{u}(\omega) \rangle &=& \delta_{\mu \nu} N_{\mu} - 
  {N_{\mu} N_{\nu} \over N(1 - i \omega \tau_{d})} + {(2-\beta)N_{\mu} 
  \over \beta N (1 - i \omega \tau_{d})} 
  \left({N_{\nu} 
  (1 - 2 i \omega \tau_{d}) \over N (1 - i \omega \tau_{d})^2} - 
  \delta_{\mu\nu} \right) + {\cal O}(N^{-1}),
\end{eqnarray}
where $N=N_1+N_2$ and $\tau_{d} = (h/N) \langle dn/d\varepsilon \rangle$
is the dwell time.
The symmetry index $\beta=1$ ($2$) in the absence (presence) of a
time-reversal-symmetry breaking magnetic field; $\beta=4$ in zero magnetic
field with strong spin-orbit scattering.
Since fluctuations in $G_{\mu\nu}^{u}(\omega)$ are of relative order 
$N^{-2}$, we may directly substitute the result (\ref{eq:Tabavg}) 
into Eq.\ (\ref{eq:GIab1a}), to obtain the first two terms in the 
large-$N$ expansion of the screened admittance 
$\langle G_{\mu\nu}(\omega) \rangle$,
\begin{eqnarray} \label{eq:Gadmavg}
  \langle G_{\mu\nu}(\omega) \rangle &=& 
  \delta_{\mu \nu} N_{\mu} - {N_{\mu} N_{\nu} 
  \over N(1 - i \omega \tau)} + {(2-\beta) N_{\mu} 
  \over \beta N (1 - i \omega \tau_{d})}
  \left({N_{\nu} (1 - 2 i \omega \tau) \over 
  N (1 - i \omega \tau)^2} - \delta_{\mu\nu} \right) 
  + {\cal O}(N^{-1}),
\end{eqnarray}
where $\tau^{-1} = \tau_{d}^{-1} + 2 e^2 N/h C$ is the 
$RC$ time.
The ${\cal O}(N)$ term in the r.h.s.\ of Eq.\ (\ref{eq:Gadmavg})
is the classical admittance, the $\beta$-dependent ${\cal O}(1)$ term
is the weak-localization correction.
Notice the almost complete formal similarity between the fully screened 
result (\ref{eq:Gadmavg}) and the unscreened result (\ref{eq:Tabavg}):
Up to one term, screening results in the replacement of the dwell time
$\tau_{d}$ by the $RC$-time $\tau$. 
The fact that the similarity is not complete is the key result
of this work which we discuss below in more detail.  

The first two terms in the small-$\omega$ expansion of 
$\langle G_{\mu \nu}(\omega) \rangle$ yield the average 
d.c.\ conductance $\langle G_{\mu \nu} \rangle$ and 
emittance $\langle E_{\mu \nu} \rangle$,
\begin{mathletters}
\begin{eqnarray}
 \langle G_{\mu \nu} \rangle &=& N_{\mu} \left( \delta_{\mu \nu} - {N_{\nu} 
 / N} \right)
     + (2-\beta) ({N_{\mu}/ \beta N})
       \left( {N_{\nu} / N} - \delta_{\mu \nu} \right), 
       \label{eq:Gdcavg} \\
 \langle E_{\mu\nu} \rangle &=&
   {N_{\mu} N_{\nu} \tau / N} 
     - (2-\beta)({N_{\mu} \tau_{d} / \beta N})
       \left( {N_{\nu} / N} - \delta_{\mu \nu} \right). \label{eq:Eavg}
\end{eqnarray}
\end{mathletters}%
Eq.\ (\ref{eq:Gdcavg}) was previously 
obtained in Ref.\ \onlinecite{BarangerMello}.
For $C \to 0$, the $RC$-time $\tau$ vanishes. 
For $\beta = 2$ we then find $\langle E_{\mu \nu} \rangle = 0$, 
for $\beta=1$ the weak-localization contribution
$\langle E_{11} \rangle = - \langle E_{12} \rangle =
N_1 N_2 \langle dn/d\varepsilon \rangle/N^3 h$
leads to a positive emittance, 
while for $\beta=4$ the emittance is negative.
For comparison we mention that for complete screening,
a ballistic conductor has 
an inductive emittance $E = -(1/4h) \langle dn/d\epsilon \rangle$, 
whereas a metallic diffusive conductor behaves capacitively as
$E = (1/6h) \langle dn/d\epsilon \rangle$ \cite{CHR}. 

For simplicity, we restrict our presentation of the admittance 
fluctuations to the d.c.\ conductance $G_{\mu \nu}$ and the 
emittance $E_{\mu \nu}$ at zero temperature. 
As before, we use the diagrammatic technique of 
Ref.\ \onlinecite{BrouwerBeenakker1996a}. 
The leading $\omega$-behavior of the admittance fluctuations is 
determined by the cross-correlator 
$\mbox{cov}\, (G_{\mu \nu}, E_{\mu \nu})$ 
[Recall that $\mbox{cov}\,(x,y) = 
\langle x y \rangle - \langle x \rangle \langle y \rangle$]. We find
that $\mbox{cov}\, (G_{\mu \nu}, E_{\mu \nu})$ is unaffected by the capacitive 
interaction with the gate,
\begin{eqnarray}
  \mbox{cov}\,(G_{\mu \nu}, E_{\rho \sigma}) = 
  \mbox{cov}\,(G_{\mu \nu}, E_{\rho \sigma}^{u}) &=& -{N_{\mu} N_{\nu}\, 
  \tau_{d} \over N^2} \left( {N_{\rho} \over N} - \delta_{\mu\rho} \right) 
  \left( {N_{\sigma} \over N} - \delta_{\nu\sigma}\right).
  \label{eq:GEcov}
\end{eqnarray}
For the autocorrelator of the emittance we find
\begin{eqnarray} \label{eq:EcovInt}
  \mbox{cov}\, (E_{\mu\nu}, E_{\rho\sigma}) &=& {3\, N_{\mu} N_{\nu} \tau_{d}^2 
  \over 2\, N^2} \left( {N_{\rho} \over N} - \delta_{\mu\rho} \right) 
  \left( {N_{\sigma} \over N} - 
  \delta_{\nu\sigma}\right)  
  + {N_{\mu} N_{\nu} \tau^2 \over N^3} \left( \delta_{\mu\rho} N_{\sigma} 
    + \delta_{\nu\sigma} N_{\rho} \right) 
  \nonumber \\ && \mbox{}
  + {2\, N_{\mu} N_{\nu} N_{\rho} N_{\sigma} \tau^2 \over \tau_{d}^2 N^4} 
  (\tau^2 - \tau_{d}^2).
\end{eqnarray}
Eqs.\ (\ref{eq:GEcov}) and (\ref{eq:EcovInt}) are valid for $\beta=2$. 
In zero magnetic field ($\beta=1,4$), the permutation 
$\rho \leftrightarrow \sigma$ must be added; 
in the presence of spin-orbit scattering ($\beta=4$), 
Eqs.\ (\ref{eq:GEcov}) and (\ref{eq:EcovInt}) are
multiplied by $1/4$.

The relevant time scales for the low-frequency response
of a chaotic quantum dot are obtained from 
Eqs.\ (\ref{eq:Eavg}) and (\ref{eq:GEcov}).
The relevant time scale for the classical admittance
is the charge-relaxation time $\tau$,
while the weak-localization correction $\delta G_{\mu \nu}(\omega)$ 
and the admittance fluctuations 
are governed by the dwell time $\tau_{d}$.
Hence, to leading order in $\omega$, 
{\em the manifestation of quantum phase coherence on 
a.c.\ transport is unaffected by the Coulomb interactions.} 
For a macroscopic quantum dot, the density of states $dn/d\varepsilon \gg C/e^2$, 
so that the two characteristic time scales 
$\tau$ and $\tau_{d}$ differ considerably.

To explain this result, we first consider 
the weak-localization correction $\delta E_{\mu \nu}$
to the average emittance. 
A screening contribution to $\delta E_{\mu \nu}$ requires a magnetic-field 
dependent quantum interference
correction to the charge accumulated in the 
cavity. To first order in $\omega$, the (unscreened) charge accumulation 
at a point $\vec r$ in the dot due to the external potential change 
$\delta U_{\mu}(\omega)$ 
is determined by the 
injectivity $d \overline n_{\mu}(\vec r)/d \varepsilon$ and 
emissivity $d \underline n_{\mu}(\vec r)/d \varepsilon$ \cite{Buettiker1993,CHR}. 
For symmetry reasons, 
the ensemble averages
$\langle d \overline n_{\mu}(\vec r)/d \varepsilon \rangle$ and
$\langle d \underline n_{\mu}(\vec r)/d \varepsilon \rangle$
both equal $N_{\mu}/N$ times the
average local density of states $\langle d n(\vec r)/d\varepsilon \rangle$,
and have no magnetic-field dependent weak-localization correction. 
Hence weak localization affects how current is distributed into 
the different leads, but it does not lead to charging of 
the sample (to leading order in $\omega$). 
This explains why the relevant time scale is the dwell time $\tau_{d}$,
characteristic of the non-interacting system, and not 
the charge-relaxation time $\tau$.

Similarly, 
the screening correction to $\mbox{cov}\, (G_{\mu \nu}, E_{\mu \nu})$ 
requires correlations between $G_{\mu \nu}$ and the 
injectivity $d \overline n_{\rho}(\vec r)/d \varepsilon$ or 
emissivity $d \underline n_{\rho}(\vec r)/d \varepsilon$ \cite{Buettiker1993,CHR}.
For a chaotic cavity, we have
\begin{eqnarray}
  \mbox{cov}\, (G_{\mu \nu}, d \overline n_{\rho}(\vec r)/d \varepsilon) &=&
  \mbox{cov}\, (G_{\mu \nu}, d \underline n_{\rho}(\vec r)/d \varepsilon) =
  N_{\rho}\, \mbox{cov}\, (G_{\mu \nu}, d n(\vec r)/d \varepsilon)/N.
\end{eqnarray}
The correlator of the d.c.\ conductance and the local density of states 
vanishes for ideal leads, which is easily verified by computation 
of $\kappa_{ij} = \mbox{cov} \left( |S_{ij}|^2 , dn(\vec r)/d\varepsilon\right)$. 
For $\beta=2$ both $dn(\vec r)/d\varepsilon$ and the distribution of $S$ 
are invariant under multiplication of $S$ with a unitary matrix. 
It follows that $\kappa_{ij}$ is independent of $i$ and $j$,
hence $\kappa_{ij} = 0$.
For $\beta=1,4$ a similar argument holds. 
The absence of correlations between the density of states and 
the d.c.\ conductance is special for the case of ideal point contacts. 
Correlations between $G_{\mu \nu}$ and $dn(\vec r)/d\varepsilon$ 
are common for point contacts with tunnel barriers, 
when the scattering matrix has no uniform distribution.

The average and variance of the admittance of a chaotic quantum dot 
with only one point contact is obtained from our results by 
setting $N_{1} = N$, $N_{2} = 0$. Denoting the admittance 
of this system by $G(\omega) = G_{11}(\omega)$, 
we thus obtain
\begin{eqnarray}
  \langle G(\omega) \rangle &=& {-N i \omega \tau \over 1 - i \omega \tau} 
+ {(2-\beta) \omega^2 \tau^2 
\over \beta (1 - i \omega \tau_{d})(1-i\omega\tau)^2}, \ \
  \mbox{var}\, G(\omega) = 
  {4 \tau^4 \over \beta \tau_{d}^2} (i \omega)^2 + {\cal O}(\omega^3).
\end{eqnarray}
Note that for a single point contact 
(see also Ref.\ \onlinecite{GoparMelloBuettiker})
the leading contribution to the variance of the admittance is 
proportional to $\omega^{2}$. 
Since the a.c.\ response of such a system is purely capacitive, 
the absence of a linear term in $\mbox{var}\, G(\omega)$ and 
the weak localization correction $\delta G(\omega)$
agrees with our previous result that 
quantum interference corrections to the low-frequency
admittance of a two-probe quantum dot
are unaffected by the Coulomb interactions.

In conclusion, we have calculated the average and variance of the admittance 
of a chaotic quantum dot which is coupled to two electron
reservoirs via multichannel 
point contacts.
The quantum dot is capacitively coupled to a gate. 
In the universal regime of multichannel point contacts, 
phase coherent a.c.\ transport is characterized by weak 
localization and admittance fluctuations. 
The relevant time scale for the quantum-interference 
effects at low frequencies $\omega$ is the dwell time $\tau_{d}$, 
while the classical admittance depends on the $RC$ time $\tau$. 
Since these two time scales differ several orders of magnitude 
for a macroscopic quantum dot ($\tau \ll \tau_{d}$), 
this effect should be clearly visible in a measurement 
of the a.c.\ response of a chaotic quantum dot.

We would like to thank the organizers of the workshop on ``Quantum Chaos'' at 
the Institute for Theoretical Physics in Santa Barbara, where this research 
was started. This work was supported in part by the Dutch Science 
Foundation NWO/FOM, the Swiss National 
Science Foundation, and by the NSF under Grant no.\ PHY94--07194.

\end{document}